\documentclass[conference]{IEEEtran}
\IEEEoverridecommandlockouts

\usepackage{cite}
\usepackage{amsmath,amssymb,amsfonts}
\usepackage{algorithmic}
\usepackage{algorithm}
\usepackage{graphicx}
\usepackage{textcomp}
\usepackage{xcolor}
\usepackage{hyperref}
\def\BibTeX{{\rm B\kern-.05em{\sc i\kern-.025em b}\kern-.08em
    T\kern-.1667em\lower.7ex\hbox{E}\kern-.125emX}}

\begin{document}

\title{A James-Stein Estimator based Generalized OMP Algorithm for Robust Signal Recovery using Sparse Representation}

\author{\IEEEauthorblockN{\textbf{Debraj Banerjee}}
\IEEEauthorblockA{\textit{Electrical Engineering} \\
\textit{Indian Institute of Science}\\
Bangalore, India \\
debrajb@iisc.ac.in}
\and
\IEEEauthorblockN{\textbf{Amitava Chatterjee}}
\IEEEauthorblockA{\textit{Electrical lEngineering} \\
\textit{Jadavpur University}\\
Kolkata, India \\
amitava.chatterjee@ieee.org}
}

\maketitle

\begin{abstract}
In this paper, we introduce a novel algorithm named JS-gOMP, which enhances the generalized Orthogonal Matching Pursuit (gOMP) algorithm for improved noise robustness in sparse signal processing. The JS-gOMP algorithm uniquely incorporates the James-Stein estimator, optimizing the trade-off between signal recovery and noise suppression. This modification addresses the challenges posed by noise in the dictionary, a common issue in sparse representation scenarios. Comparative analyses demonstrate that JS-gOMP outperforms traditional gOMP, especially in noisy environments, offering a more effective solution for signal and image processing applications where noise presence is significant.
\end{abstract}


\section{INTRODUCTION}

The term compressed sensing, which means acquiring sparse signals at rates below the Nyquist rate, has gained notable interest recently. It aims to reconstruct sparse vectors from a limited set of linearly transformed measurements. This involves two key processes: \emph{sensing}, where a $K$-sparse signal vector \(\mathbf{x}\ \)(an $n$-dimensional vector with at most $K$ non-zero elements) is converted into $m$-dimensional measurements \(\mathbf{y}\) through matrix multiplication with dictionary \(\mathbf{\Phi}\); and \emph{reconstruction}, the recovery of original sparse signals from these measurements. The measurement relationship is mathematically represented as \(\mathbf{y} = \mathbf{\Phi} \mathbf{x}.\)

In recent years, sparse coding or sparse representation and collaborative representation have evolved as very prominent genres of theory or algorithms that have been extensively used to solve signal processing and image processing problems \cite{b11,b12,b13,b14,b15,b16}. Although originally these algorithms were proposed to solve face recognition problems, later, sparse and collaborative representation algorithms and concepts of dictionary learning have been abundantly employed to solve a variety of signal and image processing problems e.g. in human movement detection \cite{b17,b18}, in intruder detection \cite{b19,b20}, in biometrics \cite{b21,b22} and so on. In compressive sensing contexts, where \(n >  m\), the equation in question forms an under-determined system with more unknowns than observations. This situation makes an accurate reconstruction of the original input \(\mathbf{x}\) using a conventional inverse transform of \(\mathbf{\Phi}\) generally unfeasible. However, leveraging knowledge about the signal's sparsity and imposing specific conditions on \(\mathbf{\Phi}\), it's possible to reconstruct \(\mathbf{x}\) by solving the \(\ell_{1}\)-minimization problem, which is formulated as minimizing \(\|\mathbf{x}\|_1\) subject to \(\mathbf{\Phi} \mathbf{x} = \mathbf{y}\).

Recently, greedy algorithms for determining the support of \(\mathbf{x}\) have gained popularity as cost-effective alternatives to linear programming methods. Key algorithms in this area include orthogonal matching pursuit (OMP) \cite{b4}, regularized OMP (ROMP) \cite{b5}, stage-wise OMP (StOMP) \cite{b1}, subspace pursuit (SP) \cite{b6}, and compressive sampling matching pursuit (CoSaMP) \cite{b7}, each offering unique approaches to the challenge of sparse signal recovery.
Out of these the generalized OMP has been quite extensively studied and implemented \cite{b23,b24,b25}, but still, these analyses demonstrate there is enough scope for further studies and improvements.

\section{OMP ALGORITHM}
The orthogonal matching pursuit (OMP) algorithm is a greedy algorithm that solves, in an iterative, heuristic manner, the classical $\ell_0$ norm minimization problem:

\begin{equation*}
    \min_{\mathbf{x}}\|\mathbf{x}\|_0 \text{ subject to } \mathbf{y} = \mathbf{\Phi x}
\end{equation*}

In each iteration, the OMP sequentially employs four steps: (\emph{i}) sweep or identification, (\emph{ii}) update support or augmentation, (\emph{iii}) update provisional solution or estimation, and (\emph{iv}) update residual steps \cite{b9, b11, b18, b19}. In each iteration $k$, in \emph{sweep} stage, the algorithm chooses the most correlated column with the residual from the dictionary matrix or sensing matrix $\mathbf{\Phi}$ i.e. the column of $\mathbf{\Phi}$ that produces the highest inner product with the residual in the previous iteration
i.e. $\mathbf{r}^k$. In the \emph{update support}
step, the support set $\Lambda^k$ is updated by the
corresponding index $\mathbf{\Phi}_i$ of that atom which was identified from the sweep stage and it is termed as
$\Lambda^{k+1}$. In the \emph{update provisional solution} stage, we first create the updated subset sensing matrix $\mathbf{\Phi}_{\Lambda^{k+1}}$ by incorporating the new atom from $\mathbf{\Phi}$ corresponding to the latest $\mathbf{\Phi}_i$ chosen and then solve the minimization problem:
\[
\min_{\mathbf{x}} \big\|\mathbf{y}-\mathbf{\Phi}_{\Lambda^{k+1}}\mathbf{x} \big\|_2^2
\]

to obtain the updated sparse solution vector $\mathbf{x}_{\Lambda^{k+1}}$. This is followed by the update residual step, where the new residual $\mathbf{r}^{k+1}$ is calculated as: 

\[
\mathbf{r}^{k+1} = \mathbf{y}-\mathbf{\Phi}_{\Lambda^{k+1}}\mathbf{x}_{\Lambda^{k+1}}
\]

Although OMP presented an exciting method for solving the \(\ell_0\) norm problem effectively, the computational burden is quite extensive because in each iteration only one new non-zero element is added to the solution vector $\mathbf{x}$ and every time the vector $\mathbf{x}$ has to be newly solved, computing a new residual. The main sources of computational complexity are the sweep/identification and residual update steps. 


\section{gOMP algorithm}
In the generalized orthogonal matching pursuit (gOMP) algorithm\cite{b9}, Wang \emph{et.al}. attempted to reduce the computational complexity of the original OMP algorithm, modifying the sweep or identification step in each iteration. The main spirit of this algorithm is to choose multiple numbers of atoms (\(\mathbf{\Phi}_i\)) in each iteration, as an improvement over the OMP algorithm, where only a single atom is selected in each iteration.

It involves comparing correlations between columns of {$\mathbf{\Phi}$} and the modified residuals. The indices of columns with the highest $N$ correlations are selected for the estimated support set \(\mathbf{\Lambda}^{k}\). When \(N\) equals \(1\), gOMP reverts to OMP. The chosen indices define the extended support set at each iteration. The least-square solution is computed as:
\[
\hat{\mathbf{x}}_{\Lambda_k} = \underset{\mathbf{u}}{\arg\min} \big\|\mathbf{y}-\mathbf{\Phi}_{\Lambda_k}\mathbf{u}
\big\|_2 = \mathbf{\Phi}^\dagger_{\Lambda_k} \mathbf{y}
\]
and the residual \(\mathbf{r}^{k}\) is updated by removing the projection of \(\mathbf{y}\) from \({\mathbf{\Phi}_{\mathbf{\Lambda}^{k}}\hat{\mathbf{x}}}_{\mathbf{\Lambda}^{k}}\).
The process repeats until reaching a maximum iteration number or the residual's \(\ell_2\)-norm falls below a specific threshold.

\section{RIP BASED RECOVERY CONDITION ANALYSIS}
A crucial property of $\mathbf{\Phi}$ which ensures the exact recovery of $\mathbf{x}$ is the Restricted Isometry Property (RIP) \cite{b2}. A sensing matrix $\mathbf{\Phi}$ satisfies the RIP condition with order \(K\) if there exists a constant \(\delta \in (0,1)\) such that:

\[
\left(1 - \delta\right)\|\mathbf{x}\|_{2}^{2} \leq \|\mathbf{\Phi x}\|_{2}^{2} \leq (1 + \delta) \|\mathbf{x}\|_{2}^{2}
\]

for any $K$-sparse vector \(\mathbf{x}\) (\(\|\mathbf{x}\|_0\le K\)). For a fixed integer \(K\), the minimum value of \(\delta\) satisfying RIP is denoted by \(\delta_{K}\). It can be shown that for any $K$-sparse signal \(\mathbf{x \in}\mathbb{R}^{n}\) and the dictionary \(\mathbf{\Phi}\) satisfying the RIP condition with isometry constant \(\delta_{K+N}\ \) then gOMP algorithm will perfectly recover the sparse signal \cite{b9} (identify all atoms of $\mathbf{x}$) if:

\begin{equation*}
    \delta_{K+N} < \frac{\sqrt{N}}{\sqrt{K} + \sqrt{N}} \
\end{equation*}

\section{SIGNAL RECOVERY UNDER NOISY CONDITIONS}
Real-world signals commonly encompass inherent noise, which refers to any undesirable signal that interferes with the signal of interest. Noise can emanate from various sources, such as the heat generated within electronic components, static electrical signals in the environment, and motion between the subject and sensors. To retrieve meaningful information from the signal, it becomes imperative to eradicate the impact of noise. Consequently, the initial stages of sparse signals involve the application of filtering techniques to effectively eliminate unwanted noise and enhance the quality of the desired signal.

Due to this noise corruption, most of the sparse classification algorithms falsely identify lots of unnecessary signal components (faulty atoms from dictionary \(\mathbf{\Phi}\)) and as a result the recovered signal lost its required sparsity. This is one of the major drawbacks of the pursuit algorithms like the OMP \& gOMP algorithm. So, to denoise the measurement data along with assuring no loss in performance of sparse classification has become a great deal.

\section{PROPOSED JAMES-STEIN ESTIMATOR BASED GENERALIZED OMP ALGORITHM (JS-gOMP)}
Mathematically the spare classification problem under a noisy environment can be expressed as the following constrained optimization problem:

\[
\min_{\|\mathbf{x}\|_0\le K} \big\|\mathbf{y}-\mathbf{\Phi}\mathbf{x}
\big\|_2 \quad \text{where } \mathbf{y}=\mathbf{y}_0 + \epsilon
\]

Here \(\mathbf{y}_{0}\) is the actual measurement part and \(\mathbf{\epsilon}\) is its noise corruption. Here as per most real-life problems, we will assume the noise to be additive Gaussian white noise (AGWN) and each noise component \(\epsilon_i\) is from the Gaussian distribution \(\mathcal{N}(0, \sigma^2)\).

Now denoting the \(i_{th}\) individual atoms in the \(m \times n\) dictionary \(\mathbf{\Phi}\) as
\(\mathbf{\phi}_{i}\) we can write:
\(\mathbf{\Phi =}\left\{ \mathbf{\phi}_{i} \right\}_{i = 1}^{n}\).
So, at the identification part of the gOMP algorithm, the correlation components become:

\[
\mathbf{Cr} = \mathbf{\Phi}^T\mathbf{y} = \mathbf{\Phi}^T\left(\mathbf{y}_{0} + \epsilon\right) \Rightarrow \mathbf{C}\mathbf{r}_{i} = \mathbf{\phi}_{i}^T\left(\mathbf{y}_{0} + \epsilon\right) = \mathbf{\phi}_{i}^T\mathbf{y}_{0} + \mathbf{\phi}_{i}^T\epsilon
\]

Here \(\phi_i^T\mathbf{y}_0\) is the deterministic part corresponding to the exact measurement whereas  \(\phi_i^T\epsilon\) is the noise part corrupting the correlation data.

\begin{algorithm}[htbp]
\caption{The modified gOMP algorithm (JS-gOMP)} \label{algo:JS-gOMP}
\begin{algorithmic}[1]
    \REQUIRE Measurements $\mathbf{y}=\{\mathbf{y}_i\}_{i=1}^p\in \mathbb{R}^{m\times p}$, sensing matrix $\mathbf{\Phi}\in \mathbb{R}^{m\times n}$, sparsity $K$, and number of iterations for each section $N$ ($N \leq \min\{K, m/K\}$)
    
    \STATE \textbf{Step 1: Initialization}
    \STATE Set iteration count $k=0$, residual vector $\mathbf{r}^0 = \mathbf{y}$, mean estimator $\hat{\mathbf{\mu}}_0 = \frac{1}{p}\sum_{i=1}^p \mathbf{r}^0_i$, noise-variance estimator $\hat{\sigma}^2_0 = (p-1)^{-1}\sum_{i=1}^p (\mathbf{r}^0_i-\hat{\mathbf{\mu}}_0) \odot (\mathbf{r}^0_i-\hat{\mathbf{\mu}}_0)$, estimated support $\Lambda^0 = \emptyset$
    
    \STATE \textbf{Step 2:}
    \WHILE{$\|\mathbf{r}^k\|_2 > \epsilon$ \AND $k < \min\{K, m/N\}$}
        \STATE Compute $n$ correlations:
        \[
        \mathbf{Cr} = \mathbf{\Phi}^T \hat{\mathbf{\mu}}^k
        \]
        \STATE Use the James-Stein estimator to reduce noise in correlation:
        \[
        \widehat{\mathbf{Cr}} = \bigg(1-\frac{p-2}{\|\mathbf{Cr}\|_2^2}\hat{\sigma}_{k}^2\bigg) \odot \mathbf{Cr}
        \]
        \STATE Identify the indices of the first $N$ highest correlation magnitudes from the $n \times 1$ correlation vector $\widehat{\mathbf{Cr}}$:
        \[
        \{\phi(i)\}_{i=1}^N \, : \, |\mathbf{\Phi}_{\phi(i)}^T\hat{\mathbf{\mu}}^k| \geq |\mathbf{\Phi}_{\phi(j)}^T\hat{\mathbf{\mu}}^k| \quad \forall i\in [N]\, ,  j > N
        \]
        \[
        [N] := \{1,2,\dots, N\}
        \]
        \STATE Augment the atoms $\{\phi(i)\}_{i=1}^N$ corresponding to the $N$ highest correlations (in magnitude) with the support:
        \[
        \Lambda^{k+1} = \Lambda^k \cup \{\phi(i)\}_{i=1}^N
        \]
        \STATE Estimate the best sparse signal within support $\Lambda^{k+1}$:
        \[
        \hat{\mathbf{x}}_{\Lambda^{k+1}} = \underset{\mathbf{u}}{\arg\min} \bigg\|\mathbf{y}-\mathbf{\Phi}_{\Lambda^{k+1}}\mathbf{u} \cdot \mathbf{1}_p^T\bigg\|_2
        \]
    
        \STATE Update the residual:
        \[
        \mathbf{r}^{k+1} = \mathbf{y}-\mathbf{\Phi}_{\Lambda^{k+1}}\hat{\mathbf{x}}_{\Lambda^{k+1}} \cdot \mathbf{1}_p^T
        \]
        \STATE Estimate the new mean and variance:
        \begin{align*}
        \hat{\mathbf{\mu}}_{k+1} &= \frac{1}{p}\sum_{i=1}^p \mathbf{r}^{k+1}_i
        \\
        \hat{\sigma}^2_{k+1} &= \frac{1}{p-1}\sum_{i=1}^p (\mathbf{r}^{k+1}_i - \hat{\mathbf{\mu}}_{k+1})^2
        \end{align*}
        \STATE Update iteration count: $k = k + 1$
    \ENDWHILE
    
    \STATE \textbf{Step 3: Return}
    \[
    \hat{\mathbf{x}} = \underset{\mathbf{u}:\text{supp}(\mathbf{u})=\Lambda^k}{\arg\min} \bigg\|\mathbf{y}-\mathbf{\Phi}_{\Lambda^{k+1}}\mathbf{u} \cdot \mathbf{1}^T\bigg\|_2
    \]
\end{algorithmic}
\end{algorithm}

So, each of the noisy correlation value
\(\mathbf{C}\mathbf{r}_{i}\) resembles a random variable having mean and variance as:

\begin{align*}
\mathbb{E}\left[\mathbf{C}\mathbf{r}_{i}\right] 
&= \mathbf{\phi}_{i}^T\mathbf{y}_{0} + \mathbb{E}\left[\mathbf{\phi}_{i}^T\mathbf{\epsilon}\right]
\\
&= \mathbf{\phi}_{i}^T\mathbf{y}_{0} + \mathbf{\phi}_{i}^T\mathbb{E}\left[\mathbf{\epsilon}\right]
\\
&= \mathbf{\phi}_{i}^T\mathbf{y}_{0} + \sum_{j = 1}^{m}{\mathbf{\phi}_{ij}\mathbb{E}\left[\mathbf{\epsilon}_{j}\right]} \\
&= \mathbf{\phi}_{i}^T\mathbf{y}_{0} + \sum_{j = 1}^{m}{\mathbf{\phi}_{ij} \times 0}
\\
&= \mathbf{\mu}_{\mathbf{C}\mathbf{r}_{i}}
\end{align*}

Similarly:

\begin{align*}
\mathbb{E}\left[\left(\mathbf{C}\mathbf{r}_{i} - \mathbf{\mu}_{\mathbf{C}\mathbf{r}_{i}}\right)^{2}\right]
&= \mathbb{E}\left[\left(\mathbf{\phi}_{i}^T\mathbf{y}_{0} + \mathbf{\phi}_{i}^T\mathbf{\epsilon} - \mathbf{\phi}_{i}^T\mathbf{y}_{0}\right)^{2}\right] \\ 
&= \mathbb{E}\left[\left(\mathbf{\phi}_{i}^T\mathbf{\epsilon}\right)^{2}\right] \\
&= \mathbb{E}\left[\left(\sum_{j = 1}^{m}{\mathbf{\phi}_{ij}\mathbf{\epsilon}_{j}}\right)^{2}\right] \\ 
&= \mathbb{E}\left[\sum_{j = 1}^{m}\sum_{k = 1}^{m} \mathbf{\phi}_{ij}\mathbf{\phi}_{ik} \mathbf{\epsilon}_{j}\mathbf{\epsilon}_{k} \right] \\
&= \sum_{j \neq k} \mathbf{\phi}_{ij}\mathbf{\phi}_{ik} \mathbb{E}\left[\mathbf{\epsilon}_{j}\right]\mathbb{E}\left[\mathbf{\epsilon}_{k}\right] + \sum_{j = k} \mathbf{\phi}_{ij}^{2} \mathbb{E}\left[\mathbf{\epsilon}_{j}^{2}\right] \\
&= \sum_{j = k} \mathbf{\phi}_{ij}^{2} \sigma^2 = \sigma^2 \sum_{j = 1}^{m} \mathbf{\phi}_{ij}^{2} \\
&= \sigma^2 \|\mathbf{\phi}_{i}\|_2^2 = \sigma^2_{\mathbf{Cr}}
\end{align*}

Thus, we can see that after getting corrupted by noise all the correlation components \(\mathbf{\text{Cr}}_{i}\) are randomly distributed with mean \(\mathbf{\phi}_{i}^T\mathbf{y}_{0}\) and variance \(\sigma^{2}\). So, the correlation vector \(\mathbf{\text{Cr}}\) will have a distribution like the measurement noise as \(\mathcal{N}(\mathbf{\Phi}^T\mathbf{y}_{0},\sigma^{2}\mathbf{I})\).

Hence, if we can estimate the mean \(\mathbf{\phi}_{i}^T\mathbf{y}_{0}\) of each of the correlation elements \(\mathbf{C}\mathbf{r}_{i}\) then we can denoise the data and use the gOMP algorithm to get the optimal spare signal.

For that, we estimate mean \(\mathbf{\hat{y}}\)  and variance \({\hat{\sigma}}^{2}\) respectively as: 
\[
{\hat{\mathbf{y}}}_{0} = \frac{1}{p}\sum_{i}{\mathbf{y}_{i}}\,,\quad 
{\hat{\sigma}}^{2} = \frac{1}{p - 1}\sum_{i}{\left( \mathbf{y}_{i} - {\hat{\mathbf{y}}}_{0} \right) \odot \left( \mathbf{y}_{i} - {\hat{\mathbf{y}}}_{0} \right)} \]
where \(\odot\) is the Hadamard product. 

From the next iterations, we update the residual ensemble as 
\[\mathbf{r}^{k+1} = \mathbf{y}-\mathbf{\Phi}_{\Lambda^{k+1}}\hat{\mathbf{x}}_{\Lambda^{k+1}}\cdot \mathbf{1}_p^T \in \mathbb{R}^{m\times p}\]

\[
\big\{\mathbf{1}_p^T = [1 \,\, 1 \,\, \dots \,\, 1] \in \,\mathbb{R}^{1\times p}\big\}
\]

So, correspondingly we update our mean and variance estimators as follows:
\[
\hat{\mathbf{\mu}}_{k+1}=\frac{1}{p}\sum_{i=1}^p \mathbf{r}^{k+1}_i, \ \hat{\sigma}^2_{k+1}=\frac{1}{p-1}\sum_{i=1}^p (\mathbf{r}^{k+1}_i-\hat{\mathbf{\mu}}_{k+1})^2
\]

Now for \(p > 2\) numbers of measurements, we will use
the James-Stein Estimator (JSE) \cite{b10} as it dominates the Least-Square-Estimator (LSE) in Minimum-Mean-Square-Error (MMSE) sense
for \(p \geq 3\).

Thus before the identification of the maximum correlation components of the correlation vector \(\mathbf{Cr}=\mathbf{\Phi}^T\hat{\mu}^k\) we use the
James-Stein estimator to get a more accurate estimation of the actual correlation vector \(\mathbf{\Phi}^T\mathbf{y}_{0}\)
as:

\[\widehat{\mathbf{Cr}} = \bigg(1-\frac{p-2}{\|\mathbf{Cr} \|_2^2}\hat{\sigma}_{k}^2\bigg)\odot \mathbf{Cr}\]

\section{PERFORMANCE EVALUATION}
To observe the empirical performance of the modified gOMP algorithm (JS-gOMP), we performed computer simulations using MATLAB (R2023a). In our experiment, we employ a testing strategy to measure the effectiveness of recovery algorithms by examining the empirical frequency of exact reconstruction in noisy environments. By comparing the maximal sparsity level of the underlying sparse signals at which the perfect recovery is ensured (this point is often called critical sparsity {[}6{]}), the accuracy of the reconstruction algorithms can be compared empirically. In our simulation, the following algorithms are considered.
\begin{enumerate}
\item
  OMP algorithm.
\item
  gOMP algorithm
\item
  JS-gOMP algorithm
\end{enumerate}

In every trial, we have constructed an \(m \times n\) sensing matrix \(\mathbf{\Phi}\) with \(m=50, n=500\), wherein the matrix entries are independently drawn from a Gaussian distribution \(\mathcal{N}(0,\frac{1}{m})\). Additionally, we generate a sparse vector \(\mathbf{x}\) with $K$-sparse properties, where the support is chosen randomly. The sparse Gaussian signals are obtained from a standard normal distribution \(\mathcal{N}(0,1)\). It is crucial to emphasize that the gOMP algorithm must adhere to \(N \le K\) and \(N \le m/K\), necessitating that the value of \(N\) does not exceed \(\sqrt{m}\) (\(\because N \le \sqrt{m}\)).

In light of this constraint, we opt for values of \(N = 2, 4, 6\) in our simulations. For each recovery algorithm, we conduct multiple independent trials and plot the empirical frequency of exact reconstruction.

It is noteworthy that in the case of JS-gOMP, we have utilized an ensemble of \(p=5\) noise-corrupted \(\mathbf{y}\)-values in our analysis.

\begin{figure}[htbp]
    \centering
    \includegraphics[width=1.0\linewidth]{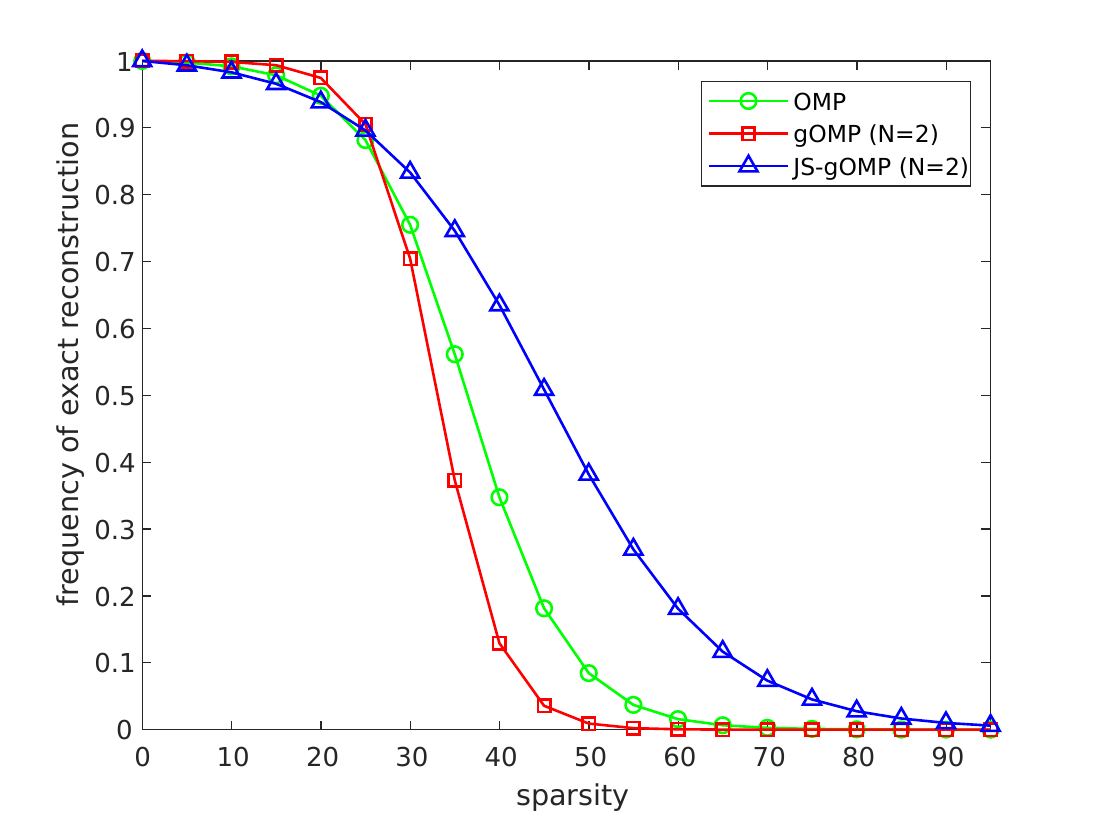}
    \caption{Reconstruction performance for $K$-sparse Gaussian signal
vectors as a function of sparsity $K$ (in a noisy environment with fixed SNR
of 4)}
    \label{fig:reconstruct}
\end{figure}

\begin{figure}[htbpt]
    \centering
    \includegraphics[width=1.0\linewidth]{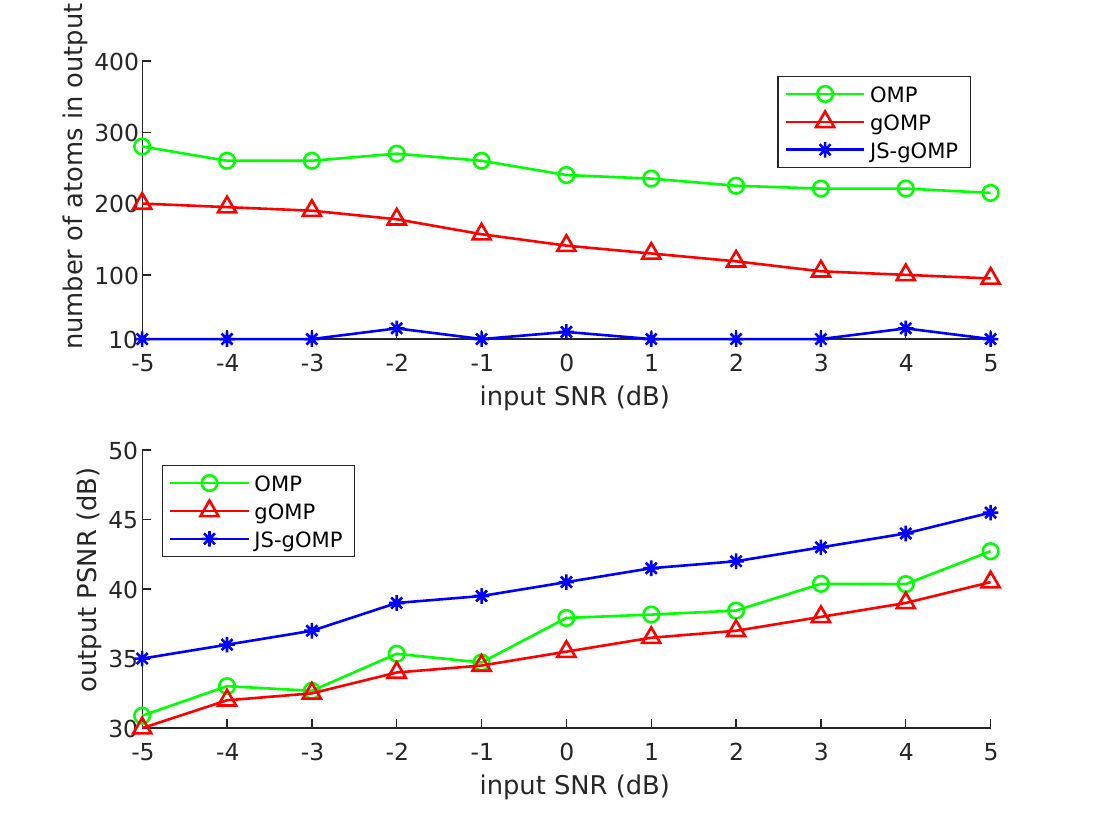}
    \caption{comparison of the number of atoms in sparse representation and corresponding PSNR values with varying noise (SNR)}
    \label{fig:atom}
\end{figure}

\begin{figure}[htbp]
    \centering
    \includegraphics[width=1.0\linewidth]{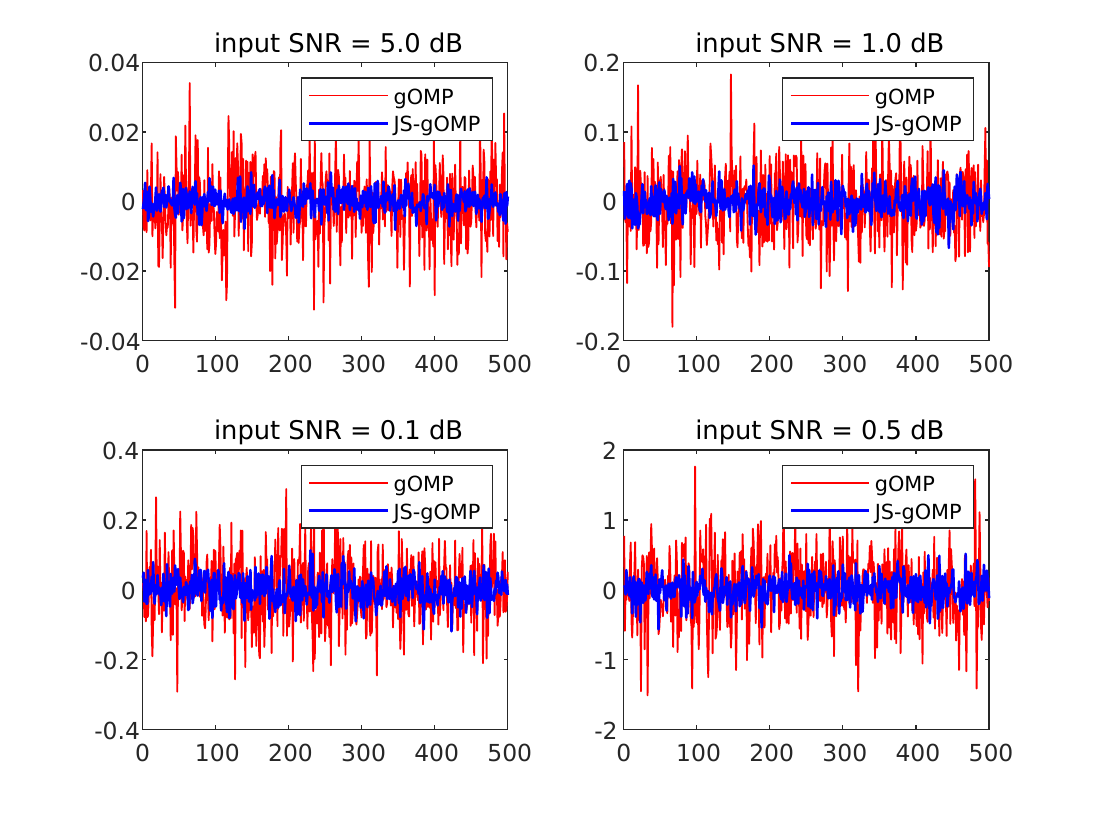}
    \caption{\centering
    reconstruction error of signal having five atoms \& $N=1$ (x-axis: input samples $n$, y-axis: exact reconstruction error i.e. $\mathbf{x}_{recon}(n)-\mathbf{x}_{actual}(n)$)}
    \label{fig:snr_plots}
\end{figure}

In Figure.\ref{fig:reconstruct}, we provide the recovery performance as a function of the sparsity level $K$. A higher level of critical sparsity indicates improved empirical reconstruction performance. The simulation results reveal that the critical sparsity of JS-gOMP algorithms is much larger compared to OMP and gOMP algorithms. As the external Gaussian noise is introduced (SNR\(\  \downarrow\)) the reconstruction performance of gOMP starts to deteriorate. But JS-gOMP gives the best results in reduced noise in the reconstructed sparse signals as per the error plots in Figure. \ref{fig:snr_plots}. As more unnecessary components start to appear in the output making the signal less sparse, the relative contribution of the actual atoms in the signal tends to reduce in both cases (OMP, gOMP), whereas the JS-gOMP gives considerably better results in reducing the noise (as per Figure. \ref{fig:atom}).

\section{CONCLUSION}
As a cost-effective solution for recovering sparse signals from compressed measurements, the OMP algorithm has received much attention in recent years. And the generalized version of the OMP algorithm i.e. gOMP is more efficient in reconstructing sparse signals. Since multiple indices can be identified with no additional post-processing operation, the proposed gOMP algorithm lends itself to parallel-wise processing, which expedites the processing of the algorithm and thereby reduces the running time. But both give poor \& erroneous results when the measurement data is corrupted by noise. In this noisy situation, our modified version of the gOMP algorithm i.e. JS-gOMP uses the James-Stine estimator which denoises the signal to a great extent and the gOMP part recovers the sparse signals with better signal-to-noise ratio.

\bibliography{biblography}

\begin{thebibliography}{10}
\providecommand{\url}[1]{#1}
\csname url@samestyle\endcsname
\providecommand{\newblock}{\relax}
\providecommand{\bibinfo}[2]{#2}
\providecommand{\BIBentrySTDinterwordspacing}{\spaceskip=0pt\relax}
\providecommand{\BIBentryALTinterwordstretchfactor}{4}
\providecommand{\BIBentryALTinterwordspacing}{\spaceskip=\fontdimen2\font plus
\BIBentryALTinterwordstretchfactor\fontdimen3\font minus \fontdimen4\font\relax}
\providecommand{\BIBforeignlanguage}[2]{{%
\expandafter\ifx\csname l@#1\endcsname\relax
\typeout{** WARNING: IEEEtran.bst: No hyphenation pattern has been}%
\typeout{** loaded for the language `#1'. Using the pattern for}%
\typeout{** the default language instead.}%
\else
\language=\csname l@#1\endcsname
\fi
#2}}
\providecommand{\BIBdecl}{\relax}
\BIBdecl

\bibitem{b11}
M.~Elad, \emph{Sparse and Redundant Representations: From Theory to Applications in Signal and Image Processing}.\hskip 1em plus 0.5em minus 0.4em\relax New York, NY, USA: Springer-Verlag, 2010.

\bibitem{b12}
J.~Wright, A.~Y. Yang, A.~Ganesh, S.~Sastry, and Y.~Ma, ``Robust face recognition via sparse representation,'' \emph{IEEE Trans. Pattern Anal. Mach. Intell.}, vol.~31, no.~2, pp. 210--227, 2009.

\bibitem{b13}
C.-Y. Lu, H.~Min, J.~Gui, L.~Zhu, and Y.-K. Lei, ``Face recognition via weighted sparse representation,'' \emph{J. Vis. Commun. Image R.}, vol.~24, no.~2, pp. 111--116, 2013.

\bibitem{b14}
L.~Zhang, M.~Yang, and X.~Feng, ``Sparse representation or collaborative representation: Which helps face recognition?'' in \emph{Proceedings of the IEEE International Conference on Computer Vision (ICCV)}, 2012, pp. 471--478.

\bibitem{b15}
C.~Zhang, H.~Li, C.~Chen, Y.~Qian, and X.~Zhou, ``Enhanced group sparse regularized nonconvex regression for face recognition,'' \emph{IEEE Trans. Pattern Anal. Mach. Intell.}, vol.~44, no.~5, pp. 2438--2452, May 2022.

\bibitem{b16}
F.~Keinert, D.~Lazzaro, and S.~Morigi, ``A robust group-sparse representation variational method with applications to face recognition,'' \emph{IEEE Trans. Image Process.}, vol.~28, no.~6, pp. 2785--2798, Jun. 2019.

\bibitem{b17}
P.~De, A.~Chatterjee, and A.~Rakshit, ``Pir sensor based aal tool for human movement detection: Modified mcp based dictionary learning approach,'' \emph{IEEE Trans. Instrum. Meas.}, vol.~69, no.~10, pp. 7377--7385, Oct. 2020.

\bibitem{b18}
------, ``Regularized k-svd-based dictionary learning approaches for pir sensor-based detection of human movement direction,'' \emph{IEEE Sens. J.}, vol.~21, no.~5, pp. 6459--6467, Mar. 2021.

\bibitem{b19}
------, ``Pir-sensor-based surveillance tool for intruder detection in secured environment: a label-consistency based modified sequential dictionary learning approach,'' \emph{IEEE Internet Things J.}, vol.~9, no.~20, pp. 20\,458--20\,466, Oct. 2022.

\bibitem{b20}
------, ``Hyperbolic function based hybrid consistent adaptive sequential dl algorithms for pir sensor based intruder detection,'' \emph{J. Instrum.}, vol.~18, no.~07, p. P07011, Jul. 2023.

\bibitem{b21}
S.~Joardar, A.~Chatterjee, and A.~Rakshit, ``A real-time palm dorsa subcutaneous vein pattern recognition system using collaborative representation-based classification,'' \emph{IEEE Trans. Instrum. Meas.}, vol.~64, no.~4, pp. 959--966, Apr. 2015.

\bibitem{b22}
S.~Joardar, A.~Chatterjee, S.~Bandyopadhyay, and U.~Maulik, ``Multi-size patch based collaborative representation for palm dorsa vein pattern recognition by enhanced ensemble learning with modified interactive artificial bee colony algorithm,'' \emph{Eng. Appl. Artif. Intell.}, vol.~60, pp. 151--163, 2017.

\bibitem{b4}
J.~A. Tropp and A.~C. Gilbert, ``Signal recovery from random measurements via orthogonal matching pursuit,'' \emph{IEEE Trans. Inf. Theory}, vol.~53, no.~12, pp. 4655--4666, Dec. 2007.

\bibitem{b5}
D.~Needell and R.~Vershynin, ``Signal recovery from incomplete and inaccurate measurements via regularized orthogonal matching pursuit,'' \emph{IEEE J. Sel. Topics Signal Process.}, vol.~4, no.~2, pp. 310--316, Apr. 2010.

\bibitem{b1}
D.~L. Donoho, I.~Drori, Y.~Tsaig, and J.~L. Starck, ``Sparse solution of underdetermined linear equations by stagewise orthogonal matching pursuit,'' \emph{IEEE Trans. Inf. Theory}, 2006, citeseer.

\bibitem{b6}
W.~Dai and O.~Milenkovic, ``Subspace pursuit for compressive sensing signal reconstruction,'' \emph{IEEE Trans. Inf. Theory}, vol.~55, no.~5, pp. 2230--2249, May 2009.

\bibitem{b7}
D.~Needell and J.~A. Tropp, ``{CoSaMP}: Iterative signal recovery from incomplete and inaccurate samples,'' \emph{Applied and Computational Harmonic Analysis}, vol.~26, no.~3, pp. 301--321, Mar. 2009.

\bibitem{b23}
R.~Qi, D.~Yang, Y.~Zhang, and H.~Li, ``On recovery of block sparse signals via block generalized orthogonal matching pursuit,'' \emph{Signal Process.}, 2018.

\bibitem{b24}
Z.~Tong, F.~Wang, C.~Hu, J.~Wang, and S.~Han, ``Preconditioned generalized orthogonal matching pursuit,'' \emph{EURASIP J. Adv. Signal Process.}, vol. 2020, no.~1, 2020.

\bibitem{b25}
Z.~Zong, T.~Fu, and X.~Yin, ``High-dimensional generalized orthogonal matching pursuit with singular value decomposition,'' \emph{IEEE Geosci. Remote Sens. Lett.}, vol.~20, pp. 1--5, 2023.

\bibitem{b9}
J.~Wang, S.~Kwon, and B.~Shim, ``Generalized orthogonal matching pursuit,'' \emph{IEEE Trans. Signal Process.}, vol.~60, no.~12, pp. 6202--6216, Dec. 2012.

\bibitem{b2}
E.~J. Candès and T.~Tao, ``Decoding by linear programming,'' \emph{IEEE Trans. Inf. Theory}, vol.~51, no.~12, pp. 4203--4215, Dec. 2005.

\bibitem{b10}
``Mse of least squares estimator (lse) vs.\ james--stein estimator (jse),'' \url{https://en.wikipedia.org/wiki/James-Stein_estimator}, [Online; accessed 24-June-2025].

\end{thebibliography}
\bibliographystyle{IEEEtran}

\end{document}